# Structure and morphology of hydroxylated nickel oxide (111) surfaces

J. Ciston, A. Subramanian, D. M. Kienzle, L.D. Marks

Department of Materials Science and Engineering, Northwestern University, 2220 Campus Drive, Cook Hall 2036, Evanston, IL, 60208-3108

## Abstract

We report an experimental and theoretical analysis of the $\sqrt{3}\times\sqrt{3}$-R30º and 2x2 reconstructions on the NiO (111) surface combining transmission electron microscopy, x-ray photoelectron spectroscopy, and reasonably accurate density functional calculations using the meta-GGA hybrid functional TPSSh. While the main focus here is on the surface structure, we also observe an unusual step morphology with terraces containing only even numbers of unit cells during annealing of the surfaces. The experimental data clearly shows that the surfaces contain significant coverage of hydroxyl terminations, and the surface structures are essentially the same as those reported on the MgO (111) surface implying an identical kinetically-limited water-driven structural transition pathway. The octapole structure can therefore be all but ruled out for single crystals of NiO annealed in or transported through humid air. . The theoretical analysis indicates, as expected, that simple density functional theory methods for such strongly-correlated oxide surfaces are marginal, while better consideration of the metal d-electrons has a large effect although, it is still not perfect.

# 1. Introduction

Polar oxide surfaces are a topic which has seen increased interest over the last few years. In terms of the formal valence, that is a fully ionic model, some rearrangement of the surface is necessary beyond a simple bulk truncation as otherwise the surface energy (enthalpy) is infinite. Of course, few oxides are as ionic as the formal valence would predict, for instance titanium in $SrTiO_3$ is closer to a true charge of $Ti^{2+}$ than the nominal charge of $Ti^{4+}$. Exceptions to this are strongly electropositive elements such as Mg or Ni where the deviation from $Mg^{2+}$ or $Ni^{2+}$ is small in the rocksalt oxide phase. When truncated along the [111] direction, rocksalt oxides exhibit alternating planes of metal$^{2+}$ and oxygen$^{2-}$ ions to create a Type-3 polar surface [1]. NiO is particularly interesting among rocksalt oxides due to its anti-ferromagnetic structure with a Neel temperature of 250°C which is useful for exchange-coupling [2] in giant magnetoresistive sensors [3].

One of the most immediately satisfying solutions to the problem of infinite dipoles at the (111) surface of rocksalt oxides is the p(2x2) reconstruction proposed by Wolf [4]. In this proposed structure, ¾ of the atoms in the first layer are missing followed by a layer in which ¼ of the bulk-like sites are vacant. This structure has the advantage of being valence neutral in a proper surface excess sum canceling the residual net dipole. Density functional theory (DFT) calculations have established that the 2x2 octapole reconstruction is the thermodynamically most stable structure over a wide range of oxygen chemical potential for the NiO (111) [5] and MgO (111) [6] surfaces, as well as over a wide range of water chemical potential on the MgO (111) surface [7]. The vast majority of experimental studies have found a reconstruction with 2x2 periodicity to be common for the (111) surface of both NiO [8-15] and MgO [7; 16; 17] which is highly suggestive. However, it is important to note that the octapole structure has never been observed on MgO (111) as an isolated phase; p(2x2) periodicity is necessary, though not sufficient proof of the octapole structure. Additionally, the attribution of NiO (111) p(2x2) structures to the octapole reconstruction has not been without controversy with different groups refining an O-terminated octapole [8] and a Ni-terminated vacancy structure [13] from the *same* set of diffraction data.

Another critical issue with oxide surfaces has been the tendency in some cases to neglect the possible presence of hydrogen in the form of hydroxide at the surface. Hydrogen is notoriously difficult to detect from surface diffraction measurements, and while it can be determined by, for instance, careful examination of a higher binding energy shoulder on the oxygen 1s peak in x-ray photoelectron spectroscopy (XPS) [7; 18], it is also easy to miss if grazing-exit geometries are not utilized. The bulk-like 1x1 surface terminations of both MgO [18] and NiO [14; 19] have both been shown experimentally to be hydroxylated. It has also been recently shown through a combination of surface diffraction, XPS, and DFT that the octapole 2x2 reconstruction is kinetically inaccessible for MgO (111) surfaces due to a water-driven structural evolution pathway wherein surface hydroxyl groups play a major role [7]. A principle goal of the present work was to determine whether this water-driven, kinetically-limited pathway was applicable to the NiO (111) system.

Surface morphology is another key factor in a number of technologically relevant processes, including thin film growth. Our present understanding of crystal morphology is based on the widely accepted theory of equilibrium faceting of crystals [20]. The basic premise of this theory is to formulate the surface free energy as a function of surface orientation and temperature, from which a Wulff construction would yield the orientation of stable facets. This simple model has been made more effective by the addition of terms to describe the various step interactions that are expected to be present in vicinal surfaces [21]. In the recent years a variety of novel surface phenomena have been reported, based on both experimental and theoretical studies [22]. These are explained by considering various types of step-step interactions, short-range attraction [23], medium-range attraction [24] and long-range attraction [25]. In addition to these interactions, the presence of a reconstruction on the terraces between steps adds another variable to the surface morphology problem. While the main focus of this paper is the atomic structure of the surface, we also report some interesting step evolution for NiO (111) surfaces annealed at various temperatures.

Nickel oxide also presents a challenge to theoretical investigations using DFT. For instance, with the simplest local-density approximation (LDA) functional (e.g. [26]), nickel oxide is a metal. With a better approach such as a generalized-gradient approximation (GGA) with the PBE formulation [27], depending upon how the calculation is done, (i.e. inclusion of spin, what pseudopotential if any are used) a DFT calculation can predict anything from a metal to a small band-gap antiferromagnetic insulator. The reason for this is that the self-interaction correction for the metal d-states is not good enough with the PBE formalism so there is too much hybridization between these states and the oxygen 2sp shell. While having an incorrect band-gap in a DFT calculation is not theoretically a problem since this is an excited-state property, not a ground-state property, the incorrect description makes nickel oxide far less ionic than reality which will have a large effect on almost all surface properties. A few studies have recently made use of more sophisticated hybrid functionals or LDA+U methods, however these have either been limited to the examination of solely the octapole structure [28; 29], or have considered only non-hydroxylated structures [5]. Furthermore, the assumption is often made that the systematic errors introduced by DFT will cancel out when relative energies are compared, but this may be wishful thinking with little justification. A well known example of this failure is the adsorption energies of CO or NO on nickel oxide surfaces or nickel atoms at the surface of $Mg_{1-x}Ni_xO$ [30] and NiO [31; 32].

In this paper, we apply a combined experimental and theoretical analysis of both the Rt3 and 2x2 reconstructions on the NiO (111) surface combining transmission electron microscopy, x-ray photoelectron spectroscopy, and reasonably accurate density functional calculations using the meta-GGA hybrid functional TPSSh. The experimental data clearly shows that both surfaces contain significant coverage of hydroxyl terminations, and the surface structures and transformation pathways are essentially the same as those on the MgO (111) surface [7]. The p(2x2) octapole structure is not observed in our experiments, and the analysis strongly suggests that this structure may be kinetically inaccessible under non-UHV annealing conditions. The theoretical analysis indicates, as expected, that simple DFT methods for such strongly-correlated oxide surfaces are marginal, while better consideration of the metal d-electrons has a large effect.

## 2. Methods and Materials

NiO single crystals used in this study were obtained from two sources: custom-grown samples of very high purity produced in a novel electric-arc furnace [33], and samples purchased commercially from MerkeTech International (99.95% pure, EPI polished 1 side). Transmission electron microscopy (TEM) is the primary tool employed in this work, primarily because of its ability to provide structural and morphological information at the atomic scale. Scanning Electron Microscopy (SEM) was also used to substantiate TEM results. Discs of 3 mm diameter cut from the parent crystal were dimpled and thinned with 5 keV $Ar^+$ ions to electron transparency. The samples were subsequently annealed in a tube furnace under flowing $O_2$ at temperatures in the range of 950°C – 1200°C to recover the surface from ion-beam damage [34] and to obtain various surface reconstructions.

Annealed NiO samples were introduced into a custom designed UHV surface science system [35] and studied using a Hitachi UHV-H9000 300 keV TEM. Structural characterization was performed using off-zone-axis transmission high energy electron diffraction (THEED) patterns. Conventional bright-field and dark-field imaging was also used to study the morphology of the annealed samples. Diffraction data was recorded on silver halide film with exposure times ranging from 2-90s and digitized to 8 bit precision and 25μm pixel pitch using an Optronix P-1000 microdensitometer calibrated to be linear. Intensities were measured using a cross-correlation technique in EDM software [36; 37]. For the Rt3 structure, a total of 700 measurements were reduced to 19 p3m1 symmetry unique reflections (p6mm Patterson symmetry) using a Tukey-biweight method to a resolution of 2 $Å^{-1}$, and similarly for the 2x2 structure, 537 measurements were reduced to 21 p3m1 symmetry unique reflections to $2Å^{-1}$. XPS data was collected according to the procedure described in [7].

## 2.1 Choice of DFT Method

While DFT calculations to accompany experimental surface structure analyses have become common, for NiO one needs to be very careful with the choice of functional. The failings of both LDA (e.g. [26]) and GGA's such as the PBE [27] are very well known for NiO. The classic method for correcting this is the LDA+U method [38; 39], which increases the ionicity of the Ni-O bond. The method requires a number for the value of the Hubbard U that is difficult to determine independently and will depend upon the local environment. A recently developed alternative which involves less in the way of arbitrary parameters, is to use an on-site hybrid [40] based upon an approach such as the PBE0 functional [41; 42]. One adds a small component of exact-exchange (which we will refer to as $\alpha$) for the relevant Kohn-Sham orbitals (d-electrons only here), which can be calculated rather simply within the muffin tin of an APW method. This is similar [40] to a LDA/GGA+U method, but with a U value that will vary with environment. A method for calculating the forces has recently been developed [43] so the on-site method can be applied for a full structural minimization with at most a 10% overhead relative to a conventional GGA

While the on-site hybrid approach appears to be better for energies (see below), we can go a bit further. The PBE functional can give poor surface energies due to problems with the long-range component decaying into the vacuum [44-46]. A much better method for this is the TPSS meta-GGA functional [47] which includes beside the gradient of the density also the kinetic energy-density in the functional form which is known to match quite well the long-range jellium surface energies. TPSS also gives much better atomization energies for molecules. It corrects, for instance, the overestimation of the atomization energy for $O_2$ which is ~6.5eV with PBE whereas TPSS gives ~5.3eV, a value that is closer to the experimental result of 5.12 eV (e.g. [48]). It is common practice with TPSS to use the electron density corresponding to e.g. a PBE potential and include only the exchange-correlation energy contribution from TPSS, which is known to give quite similar results [47]. A combination of full TPSS and PBE0 is called the TPSSh functional; the on-site version will be referred to as an on-site TPSSh method [49].

The appropriate amount of exact-exchange is in principle system dependent and a value of $\alpha=0.25$ is conventional for PBE0, a value of $\alpha=0.1$ for TPSSh; we have to answer first the question of whether these are appropriate since the on-site method has not been tested previously for energies and some recent work has suggested that there may be problems using hybrids for extended systems in a manner by which bulk, surface, and adsorbate properties are simultaneously satisfied [50; 51]. We also need to have an understanding of the magnitude of the errors in the energies that are calculated, since without this we cannot determine if two surfaces have the same energy within "theoretical error".

To analyze this, we calculated the energies of a number of test cases using PBE, TPSS, PBE0 and TPSSh with varying components of the exact exchange $\alpha$ for the later two. For the DFT calculations, the all-electron Wien2k code [52] with an augmented plane wave plus local orbital (APW+lo) basis set was employed. The problems that we considered were:

    a) The bond-energies of the series of nickel carbonyls $Ni(CO)_n$ n=1,4

    b) The atomization energy of gas-phase NiO

    c) The formation enthalpy of solid NiO

    d) The enthalpy of hydration of NiO to $Ni(OH)_2$

    e) The formation enthalpy of $Ni_2O_3$ from NiO, since with PBE $Ni_2O_3$ is more stable at STP than NiO

    f) The chemisorption enthalpy of CO and NO on NiO (001) at ½ monolayer coverage

For convenience we have grouped the gas-phase tests a) and b) and the solid-phase tests c)-e) separately, with the results plotted in Figure 1. Some relevant observations:

    1) While it is obvious that in almost all cases PBE (i.e. PBE0 with $\alpha=0$) is a bad approximation, perhaps a little unexpected PBE0 with $\alpha=0.25$ is almost as bad.

    2) In general TPSS is better than PBE (as expected), and TPSSh performs better with a smaller value of $\alpha$ than PBE0 which is consistent with what has been previously reported.

3) A value of α=0.125 ± 0.025 is reasonable for both PBE0 and TPSSh. If one wanted to use only PBE0 a slightly higher value or α=0.175 ± 0.025 would be reasonable. We prefer to use TPSSh because this is better for the long-range jellium contributions to the energy, and also does a better job than PBE for molecular bonding energies.

4) As expected, the NO and CO adsorption energies are overestimated as the DFT calculation is done at 0 K and entropic corrections have not been performed

In principle one could "tune" the value of α for a particular system, for instance bulk NiO (as is commonly done for LDA+U calculations) and then use this as the "best" value for NiO surfaces, but we believe that this is flawed logic. As mentioned above and determined specifically for $SrTiO_3$ (111) surfaces [53] the U value (screening) depends upon the local environment so the most appropriate method is to try and find the value of α which gives the best results (i.e. the best U values) for a range of different environments, i.e. a compromise value of α=0.125 ± 0.025 here with the TPSSh functional. We will stress that this particular value for α is for nickel, and can not be implicitly generalized to other atoms.

The calculations of the energies for these test cases are not perfect, neither will be the energies for surfaces. We need to have an estimate of the error to be able to determine anything. From Table 1 a reasonable estimate is to take σ ≈ 2<TPSSh-PBE0> with α=0.125 for both. For the case of NiO (111) surface structures presented in Table 2, this translates into σ ~ 0.1eV per 1x1 surface unit cell.

**2.2 Methods for NiO+$H_2O$ Surface DFT Calculations**

For all the surface structures described below, technical parameters for the surface structures were: atomic sphere sizes (RMT's) of 2.0, 1.2, and 0.6 a.u. for Ni, O and H respectively, a Fourier series cutoff of GMAX = 18 for the charge density and potential, and a wavefunction cutoff (defined as product of the smallest RMT times the largest K in the plane wave expansion) of RKMAX = 2.75. The Brillouin zone sampling was 5x5x1 for the 1x1 cell, scaled for the larger cells to retain approximately the same density of points in reciprocal space. A small (0.0018 Rydberg) temperature factor corresponding to the Fermi-Dirac occupation at room temperature was used; this had little effect since most of the relevant structures were insulators. The separation between the two surfaces was at least 1 nm, with total slab sizes of 2-2.5 nm. Tests indicated that the intrinsic numerical errors such as convergence as a function of reciprocal-space sampling were < 0.01eV per 1x1 unit cell, which is much smaller than the variations with different functionals as detailed above. In all cases the surface energies were determined by subtracting the appropriate energies for bulk NiO and/or a single molecule of $H_2O$. For this, bulk NiO was calculated in a larger supercell (for instance a hexagonal cell with a=[110] and c=[111] ) with matching technical parameters to minimize numerical errors, and the molecule of $H_2O$ in a orthorhombic cell of dimensions 5.767x6.825x6.825 $Å^3$. For reference, both PBE and the TPSS calculations were performed with the PBE minimized lattice parameters, while the PBE0 and TPSSh results were obtained with the refined lattice parameters for PBE0 and $\alpha=0.125$.

## 3. Results

### 3.1 Surface Morphology

As already mentioned, there were some interesting features of the surface steps during annealing of the samples which we will described in this section. It is important to emphasize that this study is different from other studies of vicinal surfaces [22-25], an unavoidable result of the standard TEM sample preparation procedure that involves the creation of a spherical depression. Thus the sample contains a range of miscut angles along all azimuthal orientations. While this gives us the ability to study a variety of vicinal surfaces in a single sample, the angle and orientation of the miscut cannot be quantified accurately.

Figures 2a and 2b show a diffraction pattern and bright field image obtained from a sample annealed at 950°C. The surface exhibits extended faceting with step bunches and the terraces exhibit a √3x√3R30° reconstruction (just Rt3 hereafter). The steps are winding, i.e. the kink density is relatively high, and the "flat" terraces are quite large, almost 200 nm wide. Another interesting feature in this image is the nature of the edge of the sample – the profile surfaces are very well ordered and consist of {110} planes. The step facets are generally parallel to these crystallographic planes and hence we expect the facets within the step bunches to be {110} planes. Also supporting this hypothesis is the fact that (110) surface is non-polar in rock-salt oxides and should consequently be of relatively low energy.

Upon annealing the same sample at a higher temperature of 1150°C the overall surface morphology does not change significantly. The surface still comprises of step bunches and wide terraces, the only difference being that the terraces now have a p(2x2) reconstruction. Figures 3a and 3b show the structure and morphology of this surface. Again the steps run along the <110> directions and the kink density is relatively high. These observations are consistent with previous studies on other oxide systems like MgO [7] and SrTiO$_3$ [34]. These systems exhibit a variety of reconstructions with periodicities that usually increase with temperature, with little change in surface morphology.

Samples were also annealed at 1050°C to study the transformation between the two surface phases, revealing a unexpected morphology. Figures 4a and 4b show the nature of the process. The diffraction pattern shows extended streaking along <110> directions, in addition to the presence of reflections due to the Rt3 reconstruction. This implies the presence of a continuous range of periodicities normal to these directions. Since it is impossible to construct a structure by modulating the surface unit cell to accommodate the various periods, the only viable option is to consider an ordering of surface steps. Indeed, a high-resolution dark field micrograph of the surface confirms this hypothesis. The micrograph suggests that the surface comprises of step bunches which are separated by "terraces" which have single or multiple-atom high steps spaced almost periodically. This can be seen more clearly in Figure 5, an SEM image from the same surface. In addition to pseudo-ordering of steps, the isolated kinks present in the steps have coalesced to form step edges that are very straight over length scales of 500 nm. This behavior is contrary to what one would expect from energetic or entropic considerations [22].

Figure 6 is a diffraction pattern taken after further annealing to 1100 $^{o}$C where the 2x2 phase can be seen alongside the parent √3 phase. The spots from the 2x2 phase do not show any streaking while the √3, 1x1 and the bulk spots display streaking (The √3 phase has spots that overlap with these spots). Hence we conclude that the surface with the novel step structure transforms to the 2x2 phase at higher temperatures.

**3.2 Direct methods**

In this section we present structures solved from direct methods analysis of THEED data for both the Rt3 and 2x2 reconstructed samples. There are two related structures suggested by the direct methods potential maps of the Rt3 reconstruction, which are actually structural Babinet solutions (i.e. one solution adds 1/3 of a layer to the surface and the other removes 1/3 of a layer from the surface). It is important to note that these two structures are fundamentally indistinguishable from in-plane diffraction data as they represent the same wave functions 180 degrees out of phase. This is a well-known problem with in-plane direct methods for surfaces [54].

Neither of these Rt3 structures are valence neutral and therefore do not cancel the excess surface dipole. However, two additional structures are possible, formed by the addition of a hydrogen atom per unit cell to the Rt3-Ni (in the second layer) or Rt3-O (as an OH termination) to produce Rt3-OH and Rt3-NiH structures. While with care hydrogen can be detected at a surface in a diffraction experiment once other bonding terms are properly accounted for [55], in general this is very difficult and thus not discernable from examination of the direct methods potentials. Direct methods potential maps for the two hydrogenated Rt3 solutions are shown in Figure 2c,d with overlays of the proposed structures. The Ni and O positions of the un-hydrogenated Rt3 structures are essentially identical to the hydrogenated structures and therefore not shown. These hydroxylated structures are identical to those reported recently at the MgO (111) surface [7] under similar annealing conditions. Neither the Rt3-NiH nor Rt3-OH structures are consistent with previous results which suggested a structure based upon cyclic ozone units [56], however the cyclic ozone structure is not valence neutral, and the proposed 1.4Å O-O bond distances would be expected to be energetically unfavorable.

Direct methods solutions of the diffraction data from the 2x2 structure yields a potential map shown in Figure 3c with features in the open hexagonal arrangement of the 2x2-α structure observed on MgO (111) [7; 16], and are inconsistent with the octapole reconstruction. As was the case in our previous MgO work, the localization of atomic features in the potential map to high-symmetry special sites, combined with the lack of out-of-plane information from electron diffraction inhibits the assignment of specific species to these sites. The oxidative annealing conditions should rule out the possibility of a Ni-terminated 2x2-α phase, but further XPS and DFT results are required to rationalize the assignment of O, OH, and vacancies to these sites.

**3.3 XPS**

XPS scans of an ex-situ annealed Rt3 TEM specimen exhibited a shoulder on the O-1s peak located 1.8eV higher in binding energy than the primary peak, which is consistent with the presence of O-H bonds at the surface. The sample was then annealed *in-situ* under UHV conditions at 300 °C for 12 hours after which the surface remained hydroxylated with no change in the relative energy of the shoulder. Further annealing of the same sample *in-situ* at 500 °C for 3 hours yielded no change in the position of the OH shoulder and the area of the OH feature remained constant to within 15% through all of the annealing steps (Figure 7). This suggests that chemisorbed OH groups rather than physisorbed water is responsible for the high energy feature observed in the XPS data.

The limited supply of high purity NiO material as described in [33] was exhausted before XPS data from the 2x2-α reconstructed surface could be collected. Subsequently purchased commercial crystals were unable to form the 2x2 reconstruction. This presumably related to significantly higher measured levels of Ca, Sr, and Ti impurities in the commercially grown sample.

**3.4 DFT Thermodynamics**

Surface energies for a wide variety of proposed structures were calculated and normalized to the 1x1 unit cell area using both the PBE0 and the TPSSh functionals as described above. All of calculated structures presented here were fully valence-compensated and avoid high energy excess dipole configurations, though non-compensated structures were also addressed. The structures considered in the computational portion of this work include the hydroxylated 1x1 (1x1-H), the Ni- and O-terminated hydroxylated Rt3 structures (Rt3-OH and Rt3-NiH) of [7], Ni- and O-terminated p(2x2) octapole structures of [4], hydroxylated p(2x2) vacancy structures of [13], and the 2x2-α structures of both dry (2x2-α-O1,2,3) and hydroxylated (2x2-α-OH1,2,3) structures of [7]. The TPSSh surface energies per 1x1 surface unit cell for all of these structures are plotted in Figure 8 against their respective water content. The full set of calculated surface energies can be found in Table 2. All of the valence compensated hydroxylated surfaces as well as the octapolar structures have relatively simple electronic structures with (DFT) band gaps very similar to that of bulk NiO.

We have also calculated the energies of $H_2O$ and $O_2$ molecules using both functionals since these are needed to assess the thermodynamics and plotted a surface phase diagram for a variety of NiO (111) structures against $H_2O$ chemical potential in Figure 9. It should be noted that such a diagram is useful only in the case where one assumes that the surface is fully equilibrated with respect to the gaseous environment, which may not always be the case [57]. A similar diagram which instead considered oxygen chemical potential was recently published [5], although that study did not consider hydroxylated surfaces. From Figure 9, it is clear that the hydroxylated 1x1 and dry octapole structures are the thermodynamically most stable phases over the range of water chemical potentials investigated. This is consistent with the findings of Zhang [5] which concluded that the octapole phase was the lowest in energy for all studied oxygen chemical potentials.

However, since neither the experimentally observed Rt3 nor 2x2-α phases are ever the thermodynamic minima (within the DFT error bars), it must be concluded that these are kinetically-limited metastable phases.

## 4. Discussion

### 4.1 Morphology

This study has identified an interesting material system to study the effect of the interactions between surface steps and reconstructions. These observations strongly suggest that there is a fundamental thermodynamic phenomenon at work which has minimal azimuthal anisotropy. One interesting hypothesis to explain the ordering of steps stems from the orientation of the √3 phase with respect to the step edges, namely <110> directions. Since the unit cell for the √3 phase is rotated by 30° from the step edge, every step edge can accommodate only half a unit cell. Hence every terrace must consist of an even number of unit cells normal to the step. Such "magic vicinals" have been predicted theoretically [58] and observed experimentally [59] only in the case of vicinal noble metal surfaces. With *a priori* knowledge of the surface structures, computational studies are underway to investigate the nature of step-step interactions to understand the driving forces behind this novel faceting transformation.

### 4.2 Atomic structures and transition pathways

As addressed earlier, the direct methods result from electron diffraction cannot distinguish between Ni and O terminated surfaces and the XPS data shows that there are hydroxyl groups at the surface, which allows us to discard the Rt3-O and Rt3-Ni structures in favor of their hydroxylated equivalents. In the DFT calculations, the Rt3-OH structure was found to be 0.26 eV/1x1 lower in energy than the Rt3-NiH configuration. Given the 0.1 eV/1x1 surface energy error estimation, this energy difference is significant to >98% indicating the confidence in the Rt3-OH structure. It should also be noted that there exists a simple transition pathway from the 1x1-H structure to the Rt3-OH structure by means of a single water desorption event per Rt3 unit cell, which requires no surface or bulk diffusion of oxygen, and retains the full cation framework of the 1x1-H structure. This is identical to the 1x1-H→Rt3-OH structural transition reported for the MgO (111) surface [7].

Turning to the p(2x2) reconstruction, the direct methods potential map unambiguously rules out the possibility of both the octapole and vacancy-model structures for the annealing conditions studied herein, and is instead consistent with a 2x2-α structural framework. We note that the energy ordering of the various hydroxylated structures on NiO is nearly identical to that previously calculated for MgO (compare Figure 9 herein with Figure 12 of reference [7]). This, along with the similar annealing profiles and diffraction data for the two systems, strongly suggests that NiO follows an identical Rt3→2x2-α-O→2x2-α-OH structural transition to that reported reported for the MgO (111) surface [7].

Although XPS data from the 2x2 reconstruction is unavailable, experimental support for the 2x2-α-O→2x2-α-OH structural transition and hydroxylation can be found in a prior study of NiO (111) single crystal surfaces by Barbier *et al* [9]. In that study, annealing NiO (111) crystals in air at 1300K for 3h (similar to our annealing conditions) produced a reconstruction with 2x2 periodicity. Further annealing in UHV at 950K incited a change in the intensity ordering of the diffraction pattern indicating a different surface structure, though the 2x2 periodicity was maintained. When the same crystal was again exposed to air at room temperature, the original 2x2 structure re-appeared. This behavior was interpreted to be indicative of a transition from a Ni-terminated octapole structure (air) to an O-terminated spinel (UHV), which were referred to as oxidized and reduced phases. This terminology of oxidized and reduced phases is quite misleading as both the octapole and spinel surfaces are stoichiometric and valence-compensated to alleviate the surface dipole problem, so annealing in air or UHV should have no oxygen-driven thermodynamic motivator for transformation. It should be noted that the 2x2-α phase was not considered as an alternative to the octapole or spinel structures in the previous study.

We offer an alternative interpretation of Barbier's results that is consistent with the desorption and re-adsorption of water to transition between the 2x2-α phases proposed by our model. To investigate the possibility of hydroxylization, Barbier introduced $H_2O$ vapor into the chamber at a pressure of $10^{-9}$ bar at room temperature (water chemical potential of -2.9 eV) after producing a 2x2 UHV phase (which we believe to be an unhydroxylated 2x2-α-O phase) and observed no change in the structure. This led to the conclusion that hydroxylization played no role in the transition between the two complimentary 2x2 phases. However, by inspection of the calculated phase diagram in Figure 9, the dry and hydroxylated 2x2-α phases do not overlap in energy at chemical potentials below -2.7eV indicating that exposure to water vapor at a chemical potential of -2.9 eV should have no affect on the dry 2x2-α-O structure of our model. Although the transition chemical potential lies only 0.2 eV lower than conditions previously probed, it is equivalent to a pressure $10^3$ higher pressures than used in that study. Alternatively, air at 50% relative humidity (0.02 atm) at room temperature represents a chemical potential of -2.5eV which is well within the range of conditions under which the dry and hydroxylated 2x2-α structures overlap in energy and fully explains the observed structural transition upon exposure to air.

## 5. Conclusions

There is an abundance of evidence to indicate that the 1x1H→Rt3-OH→2x2-α-O→2x2-α-OH structural transition pathway observed for the MgO (111) surface applies identically to the NiO (111) surface. All reconstructions are observed at similar temperatures for both surfaces, the Rt3-OH and 2x2-α have been observed simultaneously on the same crystal (Figure 6), and prior experiments support the reversible transformation between the dry and hydroxylated 2x2-α structures. The calculated phase diagrams are also nearly identical for MgO and NiO (111). This implies that the proposed transition pathway is extremely robust, as it is affected neither by complicated metal-3d/oxygen-2p hybridized bonding, nor spin coupling (NiO is anti-ferromagnetic at room temperature). This suggests that the transition pathway may be general for *all* metal oxides with rocksalt structures such as CoO, FeO, MnO, and SrO.

**Acknowledgements**

This work was supported by the National Science Foundation DMR-0455371 (JWC, DMK, and LDM), and DMR-0075834 (AKS). We are greatly indebted to Dr. Karl Merkle for providing (111) oriented NiO single crystals used in this study. The authors would also like to thank Professor Peter Blaha, and Dr. Fabien Tran for many fruitful discussions regarding hybrid DFT functionals.

**Tables**

| α | TPSSh | PBE0 | Difference |
|---|---|---|---|
| 0 | 0.71 | 0.94 | 0.23 |
| 0.05 | 0.41 | 0.73 | 0.33 |
| 0.1 | 0.23 | 0.44 | 0.21 |
| 0.125 | 0.40 | 0.25 | 0.15 |
| 0.15 | 0.68 | 0.23 | 0.45 |
| 0.175 | 0.95 | 0.29 | 0.66 |
| 0.2 | 1.22 | 0.50 | 0.72 |
| 0.225 | 1.49 | 0.74 | 0.75 |
| 0.25 | 1.76 | 0.98 | 0.78 |

Table 1: Deviation of DFT calculated energies (eV) from experimental values as a function of the amount of exact exchange added to the calculation (α) averaged over the systems presented in Figure 1.

| | TPSSh | PBE0 | # H2O / 1x1 |
|---|---|---|---|
| 1x1H | 0.00 | 0.10 | 0.5 |
| Rt3-OH | 0.79 | 0.81 | 0.167 |
| Rt3-NiH | 1.10 | 1.14 | 0.167 |
| O-Oct | 1.04 | 1.04 | 0 |
| Ni-Oct | 1.09 | 1.08 | 0 |
| 2x2-vac | 0.91 | 0.90 | 0.5 |
| 2x2-O-1 | 1.73 | 1.60 | 0 |
| 2x2-O-2 | 2.07 | 1.87 | 0 |
| 2x2-O-3 | 1.52 | 1.45 | 0 |
| 2x2-OH-1 | 1.04 | 1.00 | 0.25 |
| 2x2-OH-3 | 1.22 | 1.12 | 0.25 |
| 2x2-OH-2 | 1.15 | 1.09 | 0.25 |

Table 2: DFT calculated surface energies in eV per 1x1 unit cell area for various NiO-(111) structures using both the TPSSh and PBE0 functionals (α=0.125) with differing surface water content

**Figure Captions:**

Figure 1 Absolute error of DFT calculated formation enthalpies as a function of exact exchange contribution (α). Recommended range of α is shaded. Experimental reference data from [60-62]. Ni(OH)$_2$ errors scaled by a factor of 4 for clarity. a) PBE0 functionl, bulk materials, b) TPSSh functional, bulk materials, c) PBE0 functional, isolated molecules, d) TPSSh functional, isolated molecules e) adsorption energies of CO and NO in NiO(100)

Figure 2. a) Diffraction pattern from sample annealed at 950°C. The unit cell of √3x√3R30° reconstruction is outlined. b) Bright field image from the same surface showing step bunches and terraces. c) Rt3 direct methods potential maps overlaid with models of the Rt3-OH structure and d) Rt3-NiH structure. Ni blue, O red, H green.

Figure 3. a) Diffraction pattern from sample annealed at 1150°C. The 2x2 surface unit cell is outlined. b) Bright field image showing the surface morphology of the surface. c) 2x2 direct methods potential map overlaid with model of the 2x2-α-OH2 structure. Ni blue, O red, H green.

Figure 4 a) Diffraction pattern from sample annealed at 1050°C. Diffraction spot from the √3x√3R30° reconstruction is arrowed for reference. b) Dark field image from the surface using the (220) reflection.

Figure 5 Scanning Electron Micrograph from a TEM sample annealed at 1050°C showing the surface morphology.

Figure 6 Diffraction pattern from a sample annealed at 1100°C showing both the 2x2 and √3x√3R30° reconstructions. The 2x2 spots are marked by arrows with wider heads and √3x√3R30° spots are marked by narrow heads.

Figure 7. O-1s data and fitted Gaussians for Rt3 samples annealed ex-situ followed by additional in-situ UHV annealing at 500 °C. The feature at 526.5 eV is a charging artifact of the sample holder and should be ignored.

Figure 8 TPSSh surface energy per 1x1 unit cell at a water chemical potential of 0eV.

Figure 9 TPSSh Surface energy vs. $H_2O$ chemical potential for several trial structures with differing surface coverage of water. Computational errorbar drawn to scale. Region of chemical potential overlap between dry and hydroxylated 2x2-α structures shaded.

# References

[1]   P.W. Tasker, Journal of Physics C 12 (1979) 4977.
[2]   W.H. Meiklejohn, Journal of Applied Physics 33 (1962) 1328.
[3]   S. Soeya, S. Nakamura, T. Imagawa, S. Narishige, Journal of Applied Physics 77 (1995) 5838.
[4]   D. Wolf, Physical Review Letters 68 (1992) 3315.
[5]   W.B. Zhang, B.Y. Tang, Journal of Chemical Physics 128 (2008) 5.
[6]   W.B. Zhang, B.Y. Tang, Journal of Physical Chemistry C 112 (2008) 3327.
[7]   J. Ciston, A.K. Subramanian, L.D. Marks, Physical Review B 79 (2009) 085421.
[8]   A. Barbier, C. Mocuta, H. Kuhlenbeck, K.F. Peters, B. Richter, G. Renaud, Physical Review Letters 84 (2000) 2897.
[9]   A. Barbier, C. Mocuta, G. Renaud, Physical Review B 62 (2000) 16056.
[10]  A. Barbier, G. Renaud, Surface Science 392 (1997) L15.
[11]  A. Barbier, G. Renaud, C. Mocuta, A. Stierle, Structural investigation of the dynamics of the NiO(111) surface by GIXS, Birmingham, England, Elsevier Science Bv, 1998, 761.
[12]  D. Cappus, C. Xu, D. Ehrlich, B. Dillmann, C.A. Ventrice, K. Alshamery, H. Kuhlenbeck, H.J. Freund, Chemical Physics 177 (1993) 533.
[13]  N. Erdman, O. Warschkow, D.E. Ellis, L.D. Marks, Surface Science 470 (2000) 1.
[14]  F. Rohr, K. Wirth, J. Libuda, D. Cappus, M. Bäumer, H.J. Freund, Surface Science 315 (1994) L977.
[15]  C.A. Ventrice, T. Bertrams, H. Hannemann, A. Brodde, H. Neddermeyer, Physical Review B 49 (1994) 5773.
[16]  F. Finocchi, A. Barbier, J. Jupille, C. Noguera, Physical Review Letters 92 (2004) 136101.
[17]  R. Plass, K. Egan, C. Collazo-Davila, D. Grozea, E. Landree, L.D. Marks, M. Gajdardziska-Josifovska, Physical Review Letters 81 (1998) 4891.
[18]  V.K. Lazarov, R. Plass, H.C. Poon, D.K. Saldin, M. Weinert, S.A. Chambers, M. Gajdardziska-Josifovska, Physical Review B 71 (2005) 115434.
[19]  N. Kitakatsu, V. Maurice, C. Hinnen, P. Marcus, Surface Science 407 (1998) 36.
[20]  G. Wulff, Zeitschrift fur Krystallographie und Mineralogie 34 (1901) 449.
[21]  V.I. Marchenko, Zhurnal Eksperimentalnoi I Teoreticheskoi Fiziki 81 (1981) 1141.
[22]  H.C. Jeong, E.D. Williams, Surface Science Reports 34 (1999) 171.
[23]  V.B. Shenoy, S.W. Zhang, W.F. Saam, Physical Review Letters 81 (1998) 3475.
[24]  J. Frohn, M. Giesen, M. Poensgen, J.F. Wolf, H. Ibach, Physical Review Letters 67 (1991) 3543.
[25]  K. Arenhold, S. Surnev, P. Coenen, H.P. Bonzel, P. Wynblatt, Surface Science 417 (1998) L1160.
[26]  J.P. Perdew, Y. Wang, Physical Review B 45 (1992) 13244.
[27]  J.P. Perdew, K. Burke, M. Ernzerhof, Physical Review Letters 77 (1996) 3865.
[28]  O. Bengone, M. Alouani, J. Hugel, P. Blochl, Computational Materials Science 24 (2002) 192.
[29]  A. Wander, I.J. Bush, N.M. Harrison, Physical Review B 68 (2003) 4.

[30]     G. Pacchioni, C. Di Valentin, D. Dominguez-Ariza, F. Illas, T. Bredow, T. Kluner, V. Staemmler, Journal of Physics: Condensed Matter 16 (2004) S2497.
[31]     A. Rohrbach, J. Hafner, Physical Review B 71 (2005) 7.
[32]     A. Rohrbach, J. Hafner, G. Kresse, Physical Review B 69 (2004) 13.
[33]     K.L. Merkle, J.R. Reddy, C.L. Wiley, Materials Research Society Symposia 41 (1985) 213.
[34]     N. Erdman, L.D. Marks, Surface Science 526 (2003) 107.
[35]     C. Collazo-Davila, E. Landree, D. Grozea, G. Jayaram, R. Plass, P.C. Stair, L.D. Marks, Microscopy and Microanalysis 1 (1995) 267.
[36]     R. Kilaas, C. Own, B. Deng, K. Tsuda, W. Sinkler, L. Marks, EDM: Electron Direct Methods, Evanston, 2006.
[37]     R. Kilaas, L.D. Marks, C.S. Own, Ultramicroscopy 102 (2005) 233.
[38]     V.I. Anisimov, F. Aryasetiawan, A.I. Lichtenstein, Journal of Physics-Condensed Matter 9 (1997) 767.
[39]     V.I. Anisimov, J. Zaanen, O.K. Andersen, Physical Review B 44 (1991) 943.
[40]     F. Tran, P. Blaha, K. Schwarz, P. Novak, Physical Review B 74 (2006) 155108.
[41]     J.P. Perdew, M. Emzerhof, K. Burke, Journal of Chemical Physics 105 (1996) 9982.
[42]     C. Adamo, V. Barone, Journal of Chemical Physics 110 (1999) 6158.
[43]     F. Tran, J. Kunes, P. Novák, P. Blaha, L.D. Marks, K. Schwarz, Computer Physics Communications 179 (2008) 784.
[44]     R. Armiento, A.E. Mattsson, Physical Review B 72 (2005) 085108.
[45]     A.E. Mattsson, D.R. Jennison, Surface Science 520 (2002) L611.
[46]     J.P. Perdew, A. Ruzsinszky, G.I. Csonka, O.A. Vydrov, G.E. Scuseria, L.A. Constantin, X.L. Zhou, K. Burke, Physical Review Letters 100 (2008) 136406.
[47]     J.M. Tao, J.P. Perdew, V.N. Staroverov, G.E. Scuseria, Physical Review Letters 91 (2003) 146401.
[48]     W.O. Saxton, Journal of Microscopy-Oxford 179 (1995) 201.
[49]     V.N. Staroverov, G.E. Scuseria, J.M. Tao, J.P. Perdew, Journal of Chemical Physics 119 (2003) 12129.
[50]     M. Marsman, J. Paier, A. Stroppa, G. Kresse, Hybrid functionals applied to extended systems, Vienna, AUSTRIA, Iop Publishing Ltd, 2007.
[51]     A. Stroppa, G. Kresse, New Journal of Physics 10 (2008) 17.
[52]     K.S. P. Blaha, G. K. H. Madsen, D. Kvasnicka and J. Luitz, WIEN2k, An Augmented Plane Wave + Local Orbitals Program for Calculating Crystal Properties, Universität Wien, Austria, 2001.
[53]     L.D. Marks, A.N. Chiaamonti, B.C. Russell, M.C. Castell, F. Tran, P. Blaha, P.C. Stair, In Submission (2009).
[54]     C.J. Gilmore, L.D. Marks, D. Grozea, C. Collazo, E. Landree, R.D. Twesten, Surface Science 381 (1997) 77.
[55]     J. Ciston, L.D. Marks, R. Feidenhans'l, O. Bunk, G. Falkenberg, E.M. Lauridsen, Physical Review B 74 (2006).
[56]     M. Gajdardziska-Josifovska, R. Plass, M.A. Schofield, D.R. Giese, R. Sharma, Journal of Electron Microscopy (Tokyo) 51 (2002) S13.
[57]     O. Warschkow, Y.M. Wang, A. Subramanian, M. Asta, L.D. Marks, Physical Review Letters 100 (2008) 086102.

[58]     A. Bartolini, F. Ercolessi, E. Tosatti, Physical Review Letters 63 (1989) 872.
[59]     M. Yoon, S.G.J. Mochrie, D.M. Zehner, G.M. Watson, D. Gibbs, Physical Review B 49 (1994) 16702.
[60]     L.S. Sunderlin, D.N. Wang, R.R. Squires, Journal of the American Chemical Society 114 (1992) 2788.
[61]     R. Wichtendahl, M. Rodriguez-Rodrigo, U. Hartel, H. Kuhlenbeck, H.J. Freund, Surface Science 423 (1999) 90.
[62]     M.W. Chase (Ed.) NIST-JANAF Themochemical Tables, Fourth Edition, American Chemical Society, 1998.

**Figures:**

**Figures 1a-b**

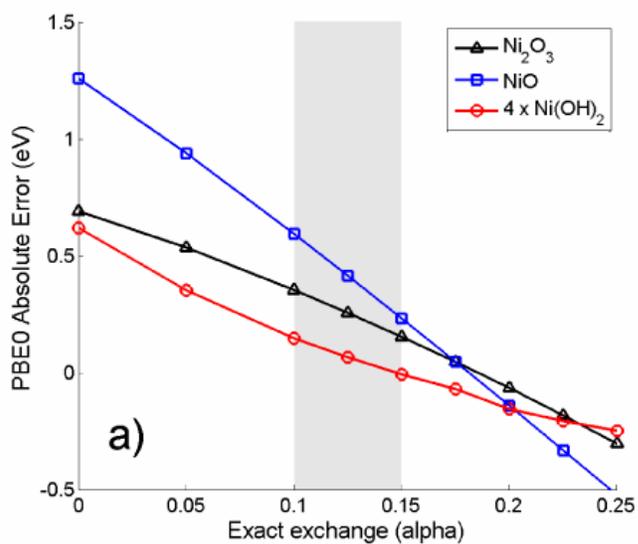

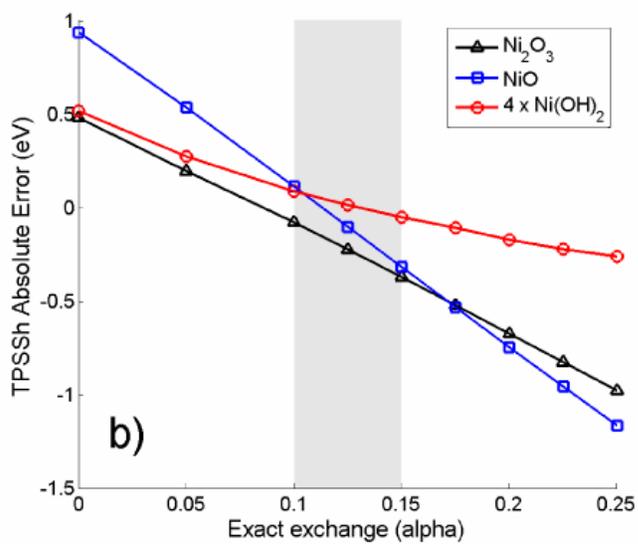

**Figures 1c-d**

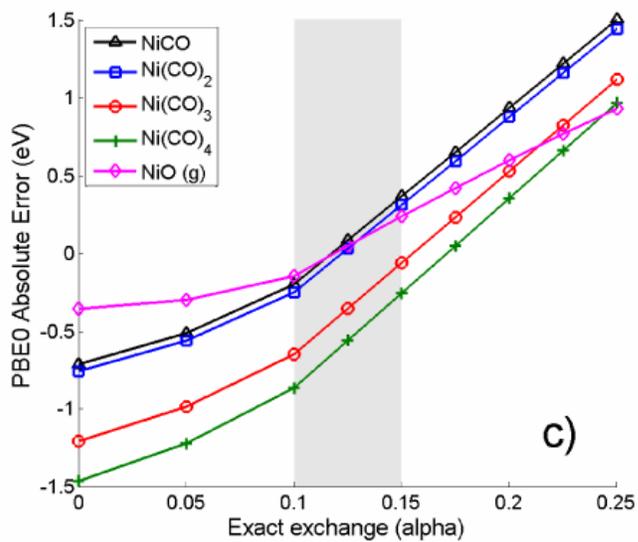

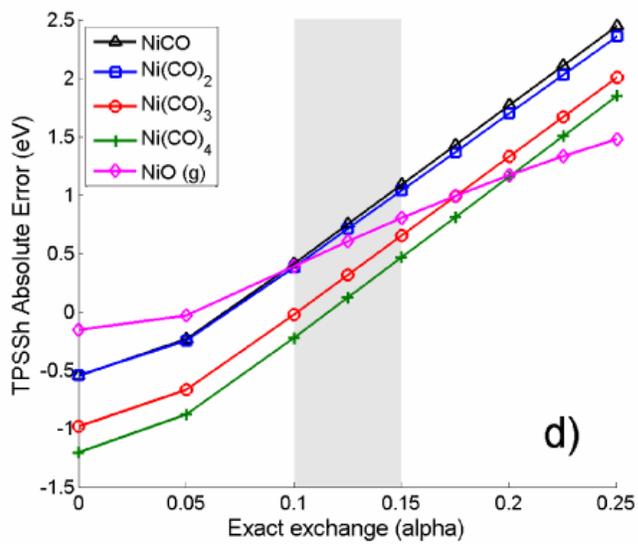

# Figure 1e

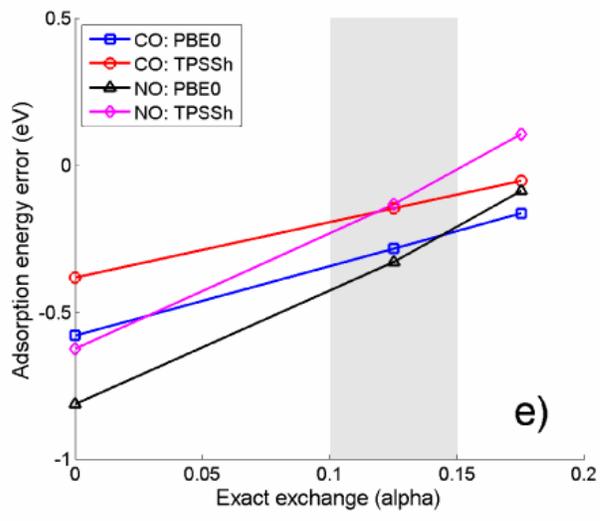

# Figures 2a and 2b

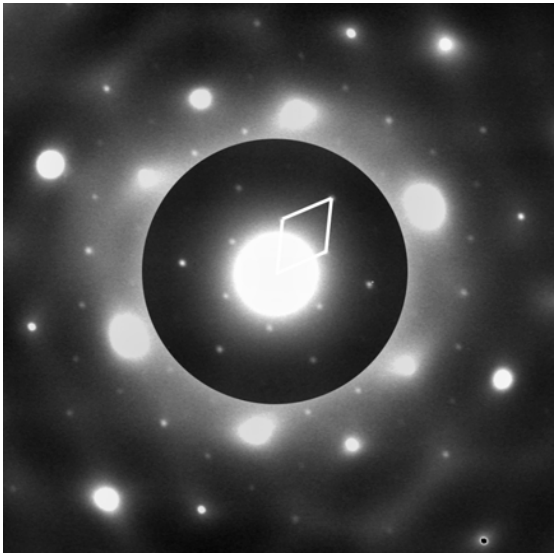
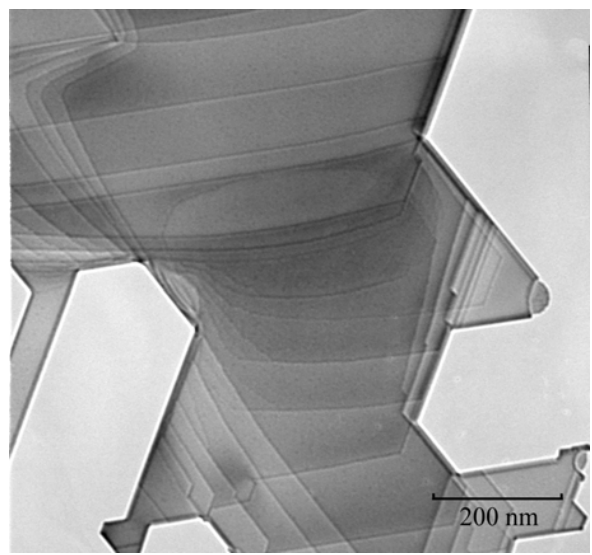

**Figures 2c and 2d**

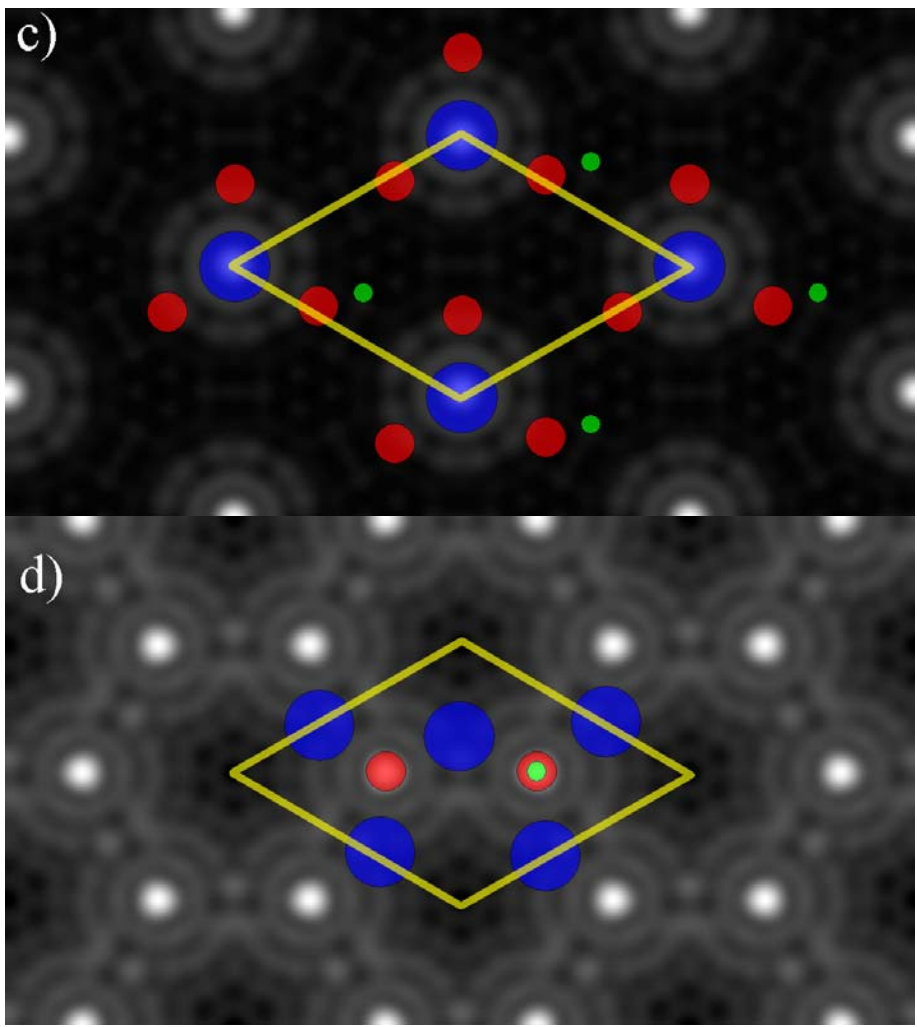

**Figures 3a-c**

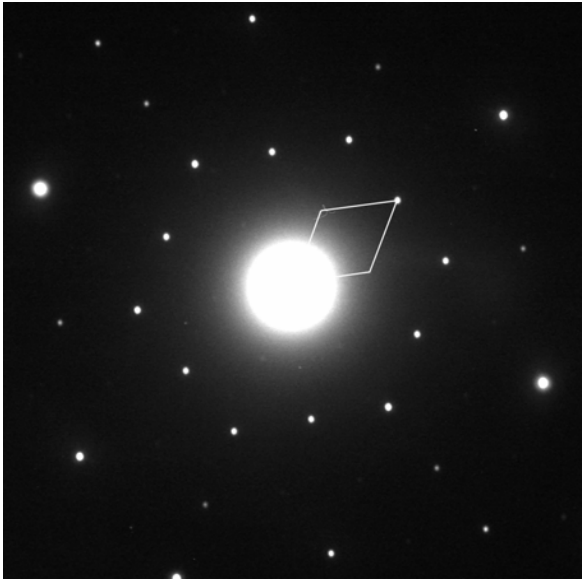
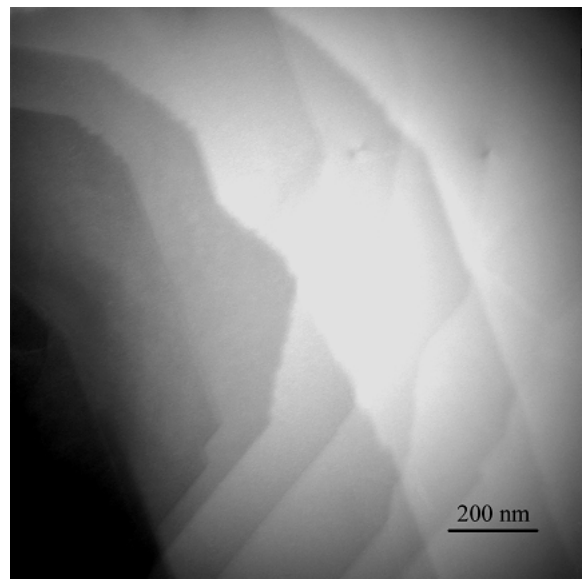
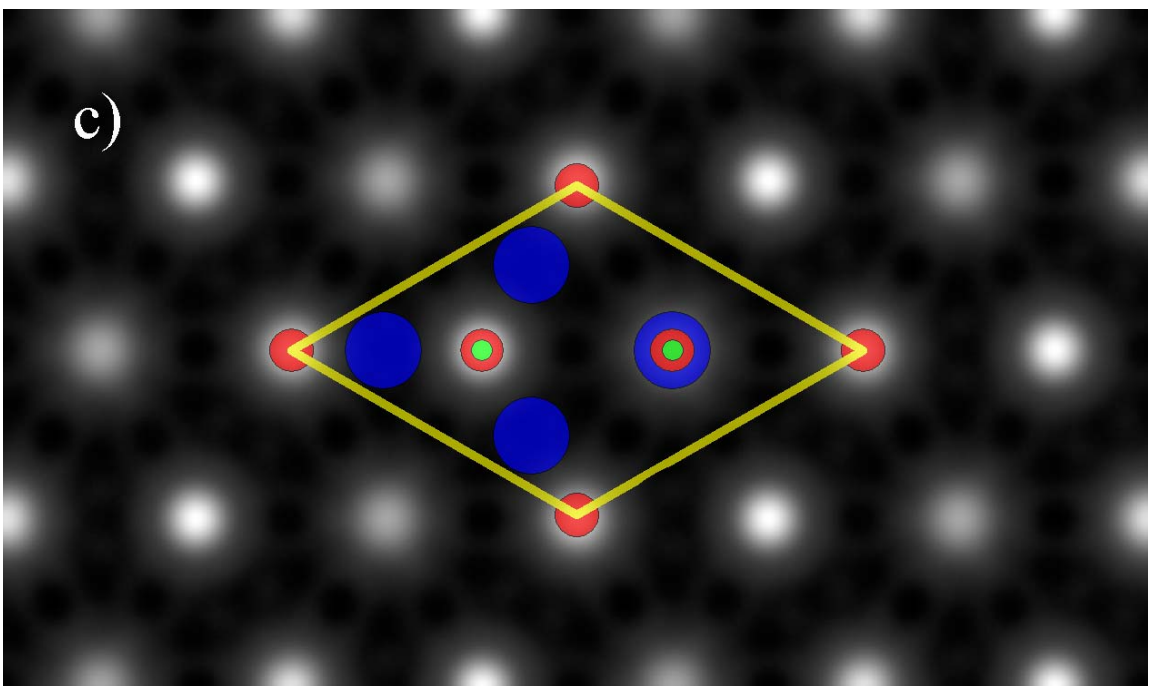

**Figures 4a and 4b**

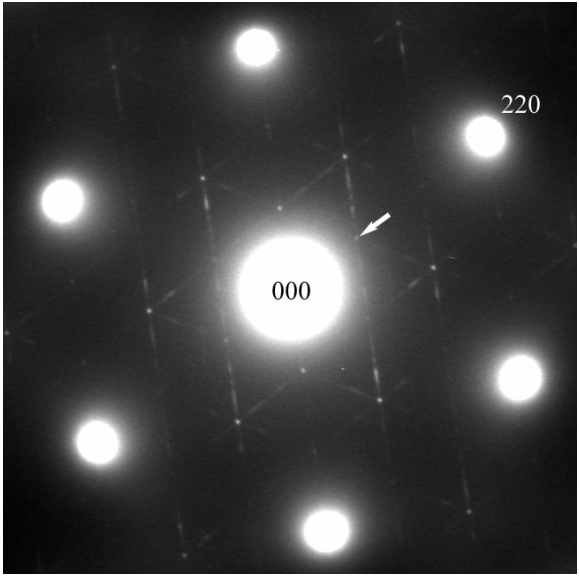

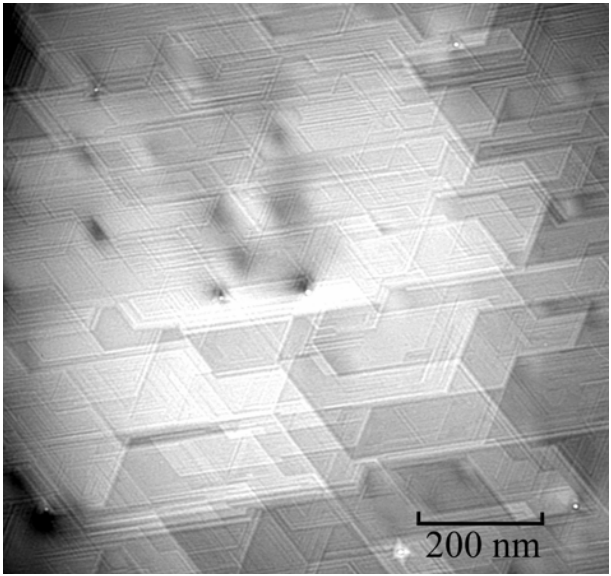

**Figure 5**

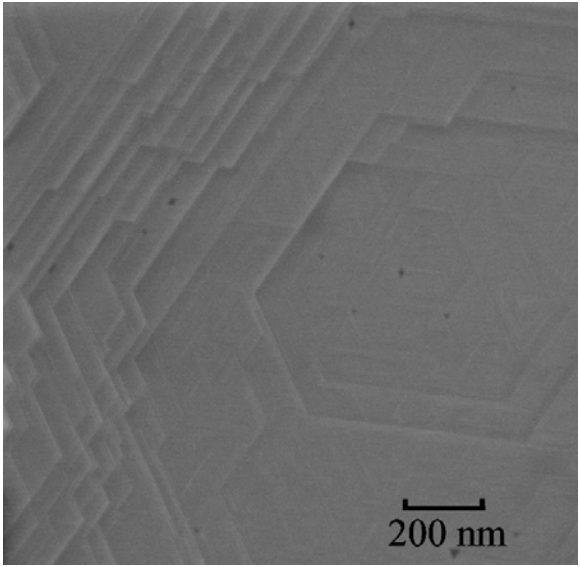

**Figure 6**

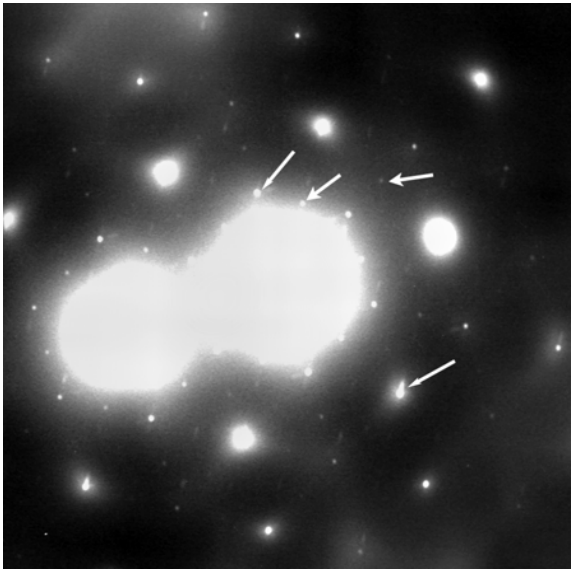

**Figure 7**

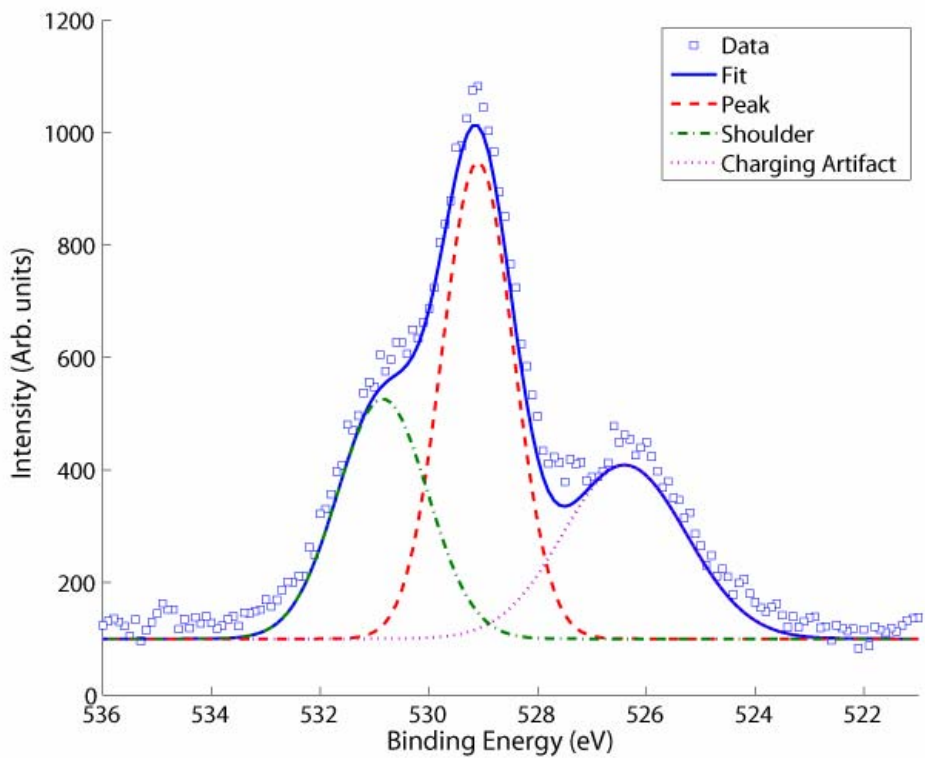

**Figure 8**

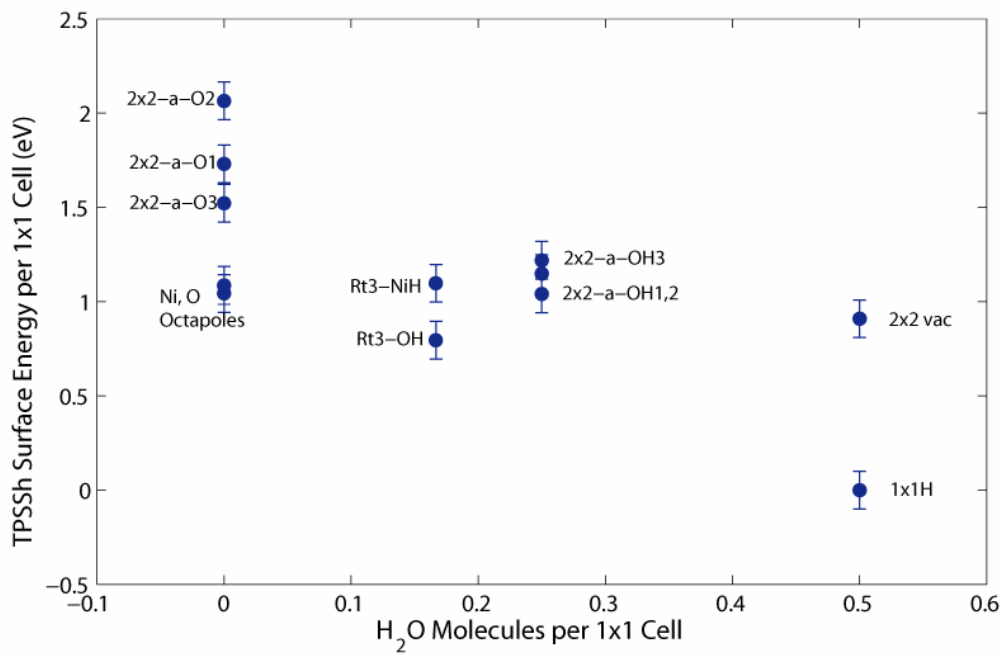

**Figure 9**

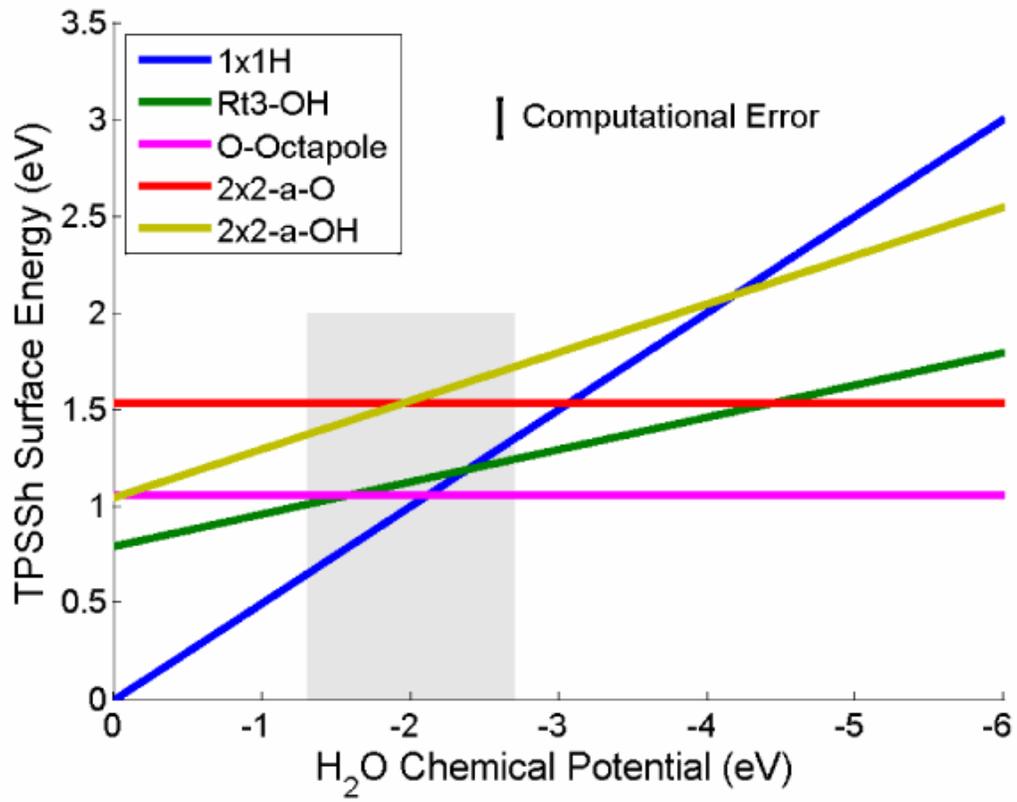